\documentstyle[12pt]{article}

\input epsf

\begin{document}
\begin{titlepage}
\today          \hfill
\begin{center}
\hfill    OITS-702 \\

\vskip .05in

{\large \bf
Can extra dimensions accessible to the SM
explain the recent measurement of anomalous
magnetic moment of the muon?}
\footnote{This work is supported by DOE Grant DE-FG03-96ER40969.}
\vskip .15in
K. Agashe \footnote{email: agashe@neutrino.uoregon.edu},
N.G. Deshpande \footnote{email: desh@oregon.uoregon.edu},
G.-H. Wu \footnote{email: gwu@electron.uoregon.edu}

\vskip .1in
{\em
Institute of Theoretical Science \\
University
of Oregon \\
Eugene OR 97403-5203}
\end{center}

\vskip .05in

\begin{abstract}

We investigate whether models with
flat extra dimensions in which SM fields propagate can give a significant 
contribution
to the anomalous
magnetic moment of the muon (MMM).
In models with only SM gauge and Higgs fields in the bulk, 
the contribution to
the MMM from
Kaluza-Klein (KK) excitations of gauge bosons is very small. This is due to 
the constraint
on the size of the extra dimensions from {\em tree-level} effects 
of KK excitations of gauge bosons on
precision electroweak observables such as
Fermi constant.
If the quarks and leptons are also allowed to propagate
in the (same) bulk (``universal'' extra dimensions), then there
are {\em no} contributions to precision electroweak observables
at tree-level. However, in this case, the constraint 
from {\em 
one-loop} contribution of KK excitations of (mainly) the top quark 
to $T$ parameter again
implies that the contribution to the MMM is small. We show that in 
models with leptons, electroweak gauge and Higgs fields
propagating in
the (same) bulk, but with quarks and gluon propagating in 
a {\em sub-space} of this bulk,
{\em both} the above constraints can be relaxed.
However, 
with only one Higgs doublet, the constraint from
the process
$b \rightarrow s \gamma$ requires
the contribution to the MMM 
to be smaller than the SM electroweak correction.
This constraint can be relaxed in models with more than one Higgs doublet.

\end{abstract}

\end{titlepage}

\newpage
\renewcommand{\thepage}{\arabic{page}}
\setcounter{page}{1}


Theories with
{\em flat} extra dimensions of size
$\sim$ (TeV)$^{-1}$
in which SM gauge (and in some cases quark and lepton)
fields propagate are motivated by 
SUSY breaking \cite{anto}, gauge coupling unification \cite{ddg},
generation of fermion
mass hierarchies \cite{as} and electroweak symmetry
breaking by a composite Higgs doublet \cite{ewsb}. 
In the effective $4D$ theory, the extra dimensions ``appear'' in the form
of Kaluza-Klein (KK) excitations of SM gauge bosons (and quarks and leptons).
These KK states have masses quantized in
units of $\sim 1 / R$, where $R$ is the size of an 
extra dimension. 
In this paper, we study the contributions to
anomalous magnetic moment of the muon (MMM)
from these KK excitations \cite{graesser, nath1, casadio}.
\footnote{It is possible that there are extra dimensions 
of size much larger than $\sim$ (TeV)$^{-1}$ 
in which gravity and 
SM singlet fields (for example,
right-handed neutrino) only propagate. In this case, 
the contribution to the MMM from 
KK excitations
of graviton (or right-handed neutrino) 
is also
important  
\cite{graesser, ng}.
Contributions to
the MMM from ``warped'' extra dimensions with and without SM gauge fields
propagating in them have also been studied \cite{rizzo1, kim, park}.}
To be specific, we investigate whether this contribution 
can be significant enough to be relevant for an explanation
of the $2.6
\sigma$ discrepancy
between the SM prediction and the new world-average value
of the MMM \cite{bnl, cm}: $a^{\hbox{exp}}_{\mu} -
a_{\mu}^{\hbox{SM}} \approx ( 43 \pm 16 ) \times 10^{-10}$. 
This discrepancy is 
(roughly) comparable to the electroweak correction to the MMM in the SM
which is $15 \times 10^{-10}$ 
\cite{cm}.
Of course, there are other new physics explanations
of this discrepancy such as SUSY, non-minimal Higgs sector
etc.

A priori, 
the one-loop
contribution to the MMM from KK excitations of electroweak gauge bosons
(and the muon) can be
comparable to the SM electroweak correction to the MMM for two reasons.
One is that the masses of these KK states (the compactification scale)
need not be much larger than the electroweak scale. The 
other is the possible
enhancement due to large number of KK states: in general, we get
\begin{equation}
a^{KK}_{\mu} \sim g_2^2 / \left( 16 \pi ^2 \right) \;
m_{\mu}^2 \; \sum_n 1 / \left( n / R \right)^2 \sim a_{\mu}^{EW}
\left( M_W R \right)^2 \; \sum_n 1 / n^2 
\label{amukk1}
\end{equation}
where $a_{\mu}^{KK}$ is the one-loop contribution to the MMM
from KK states of $W$, $Z$ and $\gamma$
and $a_{\mu}^{EW} \sim g_2^2 / \left( 16 \pi ^2 \right) \;
m_{\mu}^2 / M_W^2$ is the one-loop electroweak
correction in the SM. 
The mass of the KK state with momentum
$n_i / R$ in the $i^{\hbox{th}}$ extra dimension
is given 
by $\approx n / R$, where
$n^2 \equiv \sum_{i=1}^{i=\delta} n_i^2$ and
$n_i$ is an integer (assume $\delta$ extra dimensions
compactified on circles of radius $R$).
Thus, 
$a_{\mu}^{KK} \sim a_{\mu}^{EW}$
if $R^{-1} \sim$ few $100$ GeV and/or $\sum_n 1 / n^2 \gg 1$.
 
There are, of course, constraints on $R^{-1}$ which depend on whether
the SM quarks and leptons propagate in these extra dimensions or not.

\section{Quarks and leptons confined to 
$4D$}  

Suppose the quarks and leptons are localized 
on a ``$3$-brane'', i.e.,
do not propagate in the extra dimensions 
(the ``bulk'') in which the gauge bosons and Higgs fields 
propagate.  Then 
translation invariance in the extra dimensions is broken by
couplings of fields on
the brane to fields in the bulk -- in this case,
the couplings of quarks and leptons to gauge bosons.
In momentum space, conservation of
extra dimensional momentum is violated. Thus,
in the effective $4D$ theory
vertices with only one KK state, for example, tree-level couplings of
KK gauge bosons to leptons and quarks, 
are allowed.
These couplings 
result in
a contribution to muon decay, atomic parity violation
(APV) etc. at tree-level from Feynman diagrams
with exchange of KK $W$'s and 
$Z$'s instead of (zero-mode) $W$ and $Z$ as in
the SM \cite{graesser, nath2}.
Thus, these contributions are 
$\approx \sum_n M_W^2 / \left( n / R \right)^2$
compared to the SM. 

Then, precision electroweak measurements ($G_F$ etc.) imply an 
upper limit on 
$\sum_n M_W^2 / \left( n / R \right)^2$ of about a few percent.
and, in turn, $a^{KK}_{\mu} \ll a_{\mu}^{EW}$ (see Eq. (\ref{amukk1}))
\cite{nath1}.

\section{Universal extra dimensions}
\label{universal}

Suppose
quarks and leptons are also in the (same) bulk -- the extra dimensions
are ``universal''. 
Then, extra dimensional momentum is conserved, i.e., in the effective
$4D$ theory, ``KK number'' is conserved.
Thus, there are no vertices with only one KK
state; in particular, couplings
of (zero-modes of) quarks and leptons to KK gauge bosons are 
{\em not} allowed. 
This
implies that there are 
{\em no} 
tree-level effects on muon decay, APV, $e^+ e^- \rightarrow \mu^+ \mu^-$
etc. \cite{appel}.

However, there are constraints on
$R^{-1}$ from one loop contribution to $T$ from KK excitations of top quark,
$W$ and Higgs \cite{appel}:
\begin{eqnarray}
T & \approx & 0.83 \sum _n m_t^2 / \left( n / R \right)^2 \left[ 1 - 0.75
m_t^2 / \left( n / R \right)^2 +..\right] -
\left( 0.25 + 0.034 \delta \right) \sum _n M_W^2 / \left( n / R \right)^2
\nonumber \\
 & & - 0.043 \sum _n M_H^2 / \left( n / R \right)^2 \left[ 1 +
O \left( M_H^2 / \left( n / R \right)^2 \right) \right]
\label{Tuniv}
\end{eqnarray}
Here, $\delta$ is the number of extra dimensions and $M_H$ is the Higgs mass.
We have assumed one Higgs doublet as in the minimal SM.
The contribution of KK states of top quark
is positive and dominates over that of KK states of $W$ and Higgs
which is negative. Thus, we get a constraint
from the $2 \sigma$ {\em upper} limit $T \stackrel{<}{\sim} 0.4$ (for
$M_H \stackrel{<}{\sim} 250$ GeV). \footnote{The
SM contribution to the vacuum polarization diagrams for $W$ and
$Z$ and  
hence the $2 \sigma$ limit on $T$ depends
weakly (logarithmically) on $M_H$.}
In the above equation,  
the contribution of KK states of $W$ and Higgs at each level (i.e.,
for each value of $n$)
depends on $\delta$ and $M_H$.
In addition, the sum over KK states $\sim \sum _n 1 / n^2$ depends on $\delta$
-- in fact, $\sum_n
1 / n^2$ diverges for $\delta \geq 2$
and so has to be regulated by a cut-off.
Thus, 
the constraint on $R^{-1}$ depends on $\delta$ and $M_H$.
In any case, 
using $m_t / M_W \approx 2.1$ and
assuming $\delta \leq 6$ and $M_H \leq 250$ GeV, we get the constraint
$\left( M_W R \right)^2 \sum _n 1/ n^2 \stackrel{<}{\sim} 0.1$
(if we neglect $O \left( R / n \right)^4$ terms).
The constraint is a little bit weaker if we allow larger $\delta$ and
$M_H$. 
Nevertheless, this implies
that $a_{\mu}^{KK} \ll a_{\mu}^{EW}$.

\section{Quarks and gluon in a subspace of the bulk}

To relax the constraint from
the $T$ parameter, we need
to reduce the (positive) contribution of KK states of top quark
to $T$ compared to the (negative) contribution from KK states of
$W$ and Higgs. This can be done 
by allowing the top quark to propagate in fewer extra dimensions than
the electroweak particles. One possibility is to have only third
generation quarks stuck on a $3$-brane.  
Then, in the weak basis,
only third generation quarks couple
to KK gluons.
Thus, when we perform a unitary rotation on the
quarks to go to mass basis, there is flavor violation
at this KK gluon vertex
\cite{carone}. This results in
contribution to,  
for example, 
$B-\bar{B}$ mixing from tree-level exchange of KK gluons.
If we assume CKM-like mixing among down-type quarks, then the 
coefficient of the
$\Delta B = 2$ operator is given by
$\sim \left( V^{\star}_{tb} V_{td} \right)^2 \; g_{QCD}^2
\; \sum 1 / \left( n / R \right) ^2$.
Then, the constraint 
that this coefficient should be smaller than that in the SM
which
is $\sim \left( V^{\star}_{tb} V_{td} \right)^2 \; g_2^4 / 
\left( 16 \pi^2 \right)
\; 1 / m_t^2 \; (\hbox{or} \; 1 / M_W^2)$
implies that $\left( M_W R \right)^2 \sum_n 1 / n^2
\stackrel{<}{\sim}1/ g_{QCD}^2 \; g_2^4 / \left( 16 \pi^2 \right)
\ll 1$. 
In turn, this shows that
$a_{\mu}^{KK} \ll a_{\mu}^{EW}$.

Also, if there are some extra dimensions
in which gluons propagate but the first generation quarks do not, then
there is a tree-level contribution to $p \bar{p} \rightarrow$ jets from
exchange of KK excitations of gluons. 
This results in the constraint 
$R^{-1} \stackrel{>}{\sim} 1$ TeV from CDF \cite{cdf}: such a
large $R^{-1}$ will result in $a^{KK}_{\mu} \ll a_{\mu}^{EW}$. 

If Higgs fields are on a $3$-brane
(or in general, propagate in a sub-space of the bulk in which
electroweak gauge bosons propagate), then there is
a mass-mixing term between
zero-mode $W$, $Z$ and their KK excitations. This mixing is due to
the coupling $H^{\dagger} H W_{\mu}^{(0)}
W^{\mu \;(n)}$, where $W_{\mu}^{(n)}$ denotes a KK state of
$W$ with mass
$n/R$ and $H$ is the SM Higgs
field \cite{rizzo2, carone}. Due to this mixing, the physical
$W$ and $Z$ masses are shifted
from
the values given by the (tree-level) expression in the SM, i.e., from
$M^{SM}_W =  g v /2 $ and $M_Z^{SM}
= \sqrt{g^2 + g^{\prime \;2}} v / 2$, where $v \approx 246$ GeV
and $g$, $g^{\prime}$ are the $SU(2)$ and
$U(1)_Y$ gauge couplings. The relative shift in the
masses is given
by $\sim \sum _n M^{SM \; 2}_W \; (\hbox{or} \; M
^{SM \; 2}_Z) / \left( n / R \right)^2$,
whereas the tree-level
expression for muon decay (in terms of $v$) is the same as
in the SM since only zero-mode of $W$ contributes in this model
\cite{carone}.
As a result
precision data on $W$, $Z$ masses
give an upper limit on $\sum _n M_W^2 / \left( n / R \right)^2$ of about
a few percent.
From Eq. (\ref{amukk1}),
we see that this constraint results in $a_{\mu}^{KK}
\ll a_{\mu}^{EW}$.
So, to get $a_{\mu}^{KK} \sim a_{\mu}^{EW}$,
we have to assume that Higgs fields propagate
in the same extra dimensions as the electroweak gauge bosons.

In order to evade all these constraints
on $R^{-1}$, we assume that {\em all} quarks and gluons propagate in
the {\em same} extra dimensional space, namely, 
$\delta _c$ extra dimensions. Whereas, 
all leptons, electroweak gauge bosons {\em and} Higgs doublet
propagate in $\delta - \delta_c$ {\em additional} extra dimensions.
As mentioned before,
to evade the constraint from tree-level effects of KK states on
electroweak observables involving leptons, the leptons and electroweak
gauge bosons have to propagate in the same bulk. 
For simplicity, assume that all extra dimensions are compactified
on circles of the same
radius $R$.

A possible motivation 
for such a set-up is to explain $g_{EW}^2 \ll g_{QCD}^2$ 
at the cut-off (and hence at the scale $\sim M_Z$) as we discuss below.
The gauge coupling of the effective
$4D$ theory and the dimension{\em less} gauge coupling
of the fundamental $( 4 + \delta ) D$
theory, $g_{ ( 4 + \delta ) D}$,
are related at the cut-off $M$ 
by
$g_{4D}^2 \sim g_{ ( 4 + \delta ) D }^2 / ( M R ) ^{\delta}$ and thus if
$\delta _c < \delta$, then 
$g^2_{4D \; EW}$ is reduced (more) by extra dimensional volume than
$g^2_{4D \;QCD}$. Therefore, $g_{4D \;EW}^2 \ll g_{
4D \;QCD}^2$ at the cut-off even
though the {\em fundamental} (dimension{\em less})
gauge coupling, $g_{ ( 4 + \delta ) D}$, might be {\em same}
for QCD and electroweak gauge groups. Between
the cut-off and the energy scale $R^{-1}$, $g^2_{4D \;QCD}$ and
$g^2_{4D \;EW}$ evolve with a power-law due to the contribution
of the KK states \cite{ddg}: if $\delta _c < \delta$, then this
power-law evolution is more important for $g^2_{4D \;EW}$ than for
$g^2_{QCD}$ since there are fewer KK states with color. 
This evolution leads to an {\em additional} difference between 
$g_{4D \;QCD}^2$ and $g_{4D \;EW}^2$ at the scale $\sim M_Z$.
This is to be contrasted to
the ``standard'' picture with 
the gluon and electroweak gauge bosons propagating
in the {\em same} bulk. In this picture, due to power-law evolution
(which is {\em equally} important for both
gauge couplings), it is possible that the two {\em effective} $4D$ couplings,
i.e., $g^2_{4D \;QCD}$ and $g^2_{4D \;EW}$, 
unify 
at the cut-off $M$ (even though $M$ might not be much larger
than $R^{-1}$) \cite{ddg}. Whereas, with $\delta _c < \delta$, as 
discussed above,  
the {\em fundamental}
dimension{\em less} QCD and electroweak gauge couplings can ``unify''
at the cut-off, even though the {\em effective} $4D$ gauge couplings do not.

There are two cases within this type of model which we now discuss.
 
\subsection{Case $\delta _c = 0$: quarks and gluon on a brane}
\label{subspace1}
In this case,
there are no 
KK excitations of the top quark and hence Eq. (\ref{Tuniv}) gives
$T \approx - (0.35 - 0.8) \left( M_W R \right)^2 \sum _n 1 / n^2
\stackrel{>}{\sim} - (0.2 - 0.3)$ which is
the $2 \sigma$ lower limit on $T$:
this also depends weakly on $M_H$.
The range $0.35 - 0.8$ corresponds to varying
$\delta$ and $M_H$ (assuming $\delta \leq 6$ and $M_H \stackrel{<}{\sim}
250$
GeV).
This implies that
$\left( M_W R \right)^2 \sum _n 1/ n^2 \stackrel{<}{\sim} 1$
so that $a_{\mu}^{KK} \sim a_{\mu}^{EW}$ is barely allowed by the 
constraint
from $T$ (see Eq. (\ref{amukk1})).

However, there is a stronger constraint from inclusive hadronic decays
of $B$. The reason is that
KK states of $W$ contribute at tree-level
to hadronic weak decays, but not to
semileptonic weak decays since (zero-mode) leptons do not couple to KK
states of gauge bosons. 
Again, this additional contribution to the amplitude
is $\approx \left( M_W R \right)^2 \sum _n 1/ n^2$ 
compared to the SM. The SM prediction for the ratio of the rates
of hadronic and semileptonic inclusive $B$ decays, to be specific
$\Gamma \left( b \rightarrow c \bar{u}d \right) /
\Gamma \left( b \rightarrow c l \nu \right)$,
agrees with experiment to within $\sim 20 \%$ 
at $1 \sigma$ (combining theory
and experiment 
errors)
\cite{pdg}. Thus,  
we get $\left( M_W R \right)^2 \sum _n 1/ n^2$ 
(the ratio of KK to SM $W$ {\em amplitude})
is at most 
about $10 \%$ so that
$a_{\mu}^{KK} \ll a_{\mu}^{EW}$ 

\subsection{Case 
$\delta _c \neq 0$}
\label{subspace2}

In this case, the constraints from both
$T$ and hadronic $B$ decays
can be relaxed as we discuss below. 

Using $m_t \approx 2.1 M_W$ in Eq. (\ref{Tuniv}), we get
\begin{eqnarray}
T & \approx & \left( M_W R \right)^2 \left[ 3.9 \sum ^{\delta _c}
_n 1 / n^2 - (0.35 - 0.8) \sum ^{\delta} _n 1 / n^2 \right],
\label{T3b}
\end{eqnarray}
where (and from now on) $\sum ^{\delta^{\prime}}
_n 1 / n^2$ denotes sum over KK states of a particle with momentum in 
only $\delta^{\prime}$ of the
$\delta$ extra dimensions. As discussed
before, the
coefficient $\sim (0.35 - 0.8)$
of the $W$ and Higgs contribution to $T$ at each KK level
depends on $\delta$ and $M_H$ -- we have allowed
$\delta \leq 6$ and $M_H \leq 250$ GeV. 
It is clear that
there can be a cancelation between the contributions of top quark
and $W+$Higgs, i.e., $T \sim 0$ 
\footnote{Strictly speaking $- 0.3
\stackrel{<}{\sim} T \stackrel{<}{\sim} 0.4$ is allowed at
$2 \sigma$
(again depending weakly on $M_H$),
but for simplicity,
we assume $T \sim 0$.} if the following condition is
satisfied:
\begin{equation}
\frac{ \sum ^{\delta} _n 1 / n^2 }{ \sum ^{\delta _c}
_n 1 / n^2 } \sim (5 -10), 
\label{Tconstraint}
\end{equation}
where the RHS depends on $\delta$ and $M_H$.
In other words, this case is intermediate between
the case in section \ref{universal}, 
where the contribution from top quark
is dominant and the case in section
\ref{subspace1}, where there is only the $W$ and Higgs contribution. 

Also, in this case,
the additional contribution to hadronic $B$ decays (relative to SM) is
given by
\begin{equation}
r_B \equiv \frac{ W_{KK} \; \hbox{amplitude} }{ W \; \hbox{amplitude
in SM} } 
\approx
\left( M_W R \right) ^2 \sum
_n ^{\left( \delta - \delta _c \right)} 1 / n^2
\label{B3b}
\end{equation}
since only KK excitations of $W$ with momentum in the
$\left( \delta - \delta _c \right)$ extra dimensions in which quarks do not
propagate contribute.
From Eq. (\ref{amukk1}), to get $a_{\mu}^{KK} \sim a_{\mu}^{EW}$,
we require
\begin{equation}
\left( M_W R \right)^2 \sum _n ^{\delta} 1 / n^2 \sim 1.
\label{amuconstraint}
\end{equation}
Thus, even if
Eq. (\ref{amuconstraint}) is satisfied, the contribution
to hadronic $B$ decays can be smaller than
$\sim 10 \%$ in the amplitude (as required by the 
agreement between theory and experiment) provided 
the following condition is satisfied
\begin{equation}
\sum
_n ^{\left( \delta - \delta _c \right)} 1 / n^2
\stackrel{<}{\sim} 1/10 \; \sum^{\delta}_n 1 / n^2. 
\label{Bconstraint}
\end{equation}

Therefore, for given $\delta$,
it might be possible to evade the constraints from 
$T$ and hadronic decays (while getting 
$a_{\mu}^{KK} \sim a_{\mu}^{EW}$) by an appropriate
choice of $\delta _c$ and $M_H$ such that Eqs. (\ref{Tconstraint})
and (\ref{Bconstraint}) are satisfied. 

Next, we discuss the values of the parameters
$\delta$, $\delta _c$, $R^{-1}$ and the cut-off $M$ for which this
is possible. 

We begin with direct bounds on $R^{-1}$ from collider searches.
In this case,
there are KK excitations of quarks which appear
as heavy stable quarks at hadron colliders and
searches at CDF imply
$R^{-1} \stackrel{>}{\sim} 300$ GeV (for $\delta _c = 1$)
\cite{appel}.
The direct lower bound on $R^{-1}$ will be (slightly)
larger for
$\delta _c > 1$ since there will be more KK modes at the lowest
level, i.e., with $n = 1$.

The KK states of electroweak gauge bosons with momentum in the
$\left( \delta - \delta _c \right)$ extra dimensions (in
which quarks do not propagate) couple to
zero-mode quarks (with the same strength
as in SM), but not to zero-mode leptons. Thus, at hadron colliders,
the signatures of these KK states
will be similar to that of $W^{\prime}$ and $Z^{\prime}$ decaying to dijets.
The CDF search excludes hadronic decays of $W^{\prime}$ in the mass range
$300 < M_{W^{\prime}} < 420$ GeV at $95 \%$ confidence level \cite{cdf}.
For the KK states
of $W$, the branching ratio to quarks is slightly larger than
that for $W^{\prime}$ since the leptonic decays are absent. Also, since
there are $2 \; \left( \delta
- \delta _c \right)$ relevant states at the lowest KK level (i.e.,
with $n = 1$),
the production cross-section can
be larger than that for $W^{\prime}$ of same mass.
We will assume the limit
$R^{-1} \stackrel{>}{\sim} 400$ GeV from a combination
of the heavy stable quark and $W^{\prime}$ searches.

In turn, this implies that $\sum ^{\delta} _n 1 / n^2
\stackrel{>}{\sim} 25$ to give $a_{\mu}^{KK} \sim a_{\mu}^{EW}$
(see Eq. (\ref{amuconstraint})). Thus,
$\delta = 1, 2$ will not
suffice since $\sum _n ^{\delta} 1 / n^2 = \pi^2 / 3$ for $\delta =1$
and is log-divergent (but, $O(1)$) for $\delta =2$.
The sum
over KK states of a SM particle $\sum ^{\delta} _n 1 / n^2$ is
power-divergent if
$\delta \geq 3$. It can be approximated
by an integral (provided $M \gg R^{-1}$):
$\sum ^{\delta}_n 1 / n^2
\approx S_{\delta}
/ (\delta - 2) \;  \left( M R \right)^{\delta -2}$,
where the KK sum is cut-off at the mass scale $M$
(i.e., $n_{\hbox{max}} / R \approx M$)
and $S_{\delta} = 2 \pi^
{\delta / 2} / \Gamma[\delta / 2 ]$ (the surface
area of a unit-radius sphere in $\delta$ dimensions). 
Thus, for $\delta \geq 3$,  
we get 
\begin{equation}
\frac{ a_{\mu}^{KK} }{ a_{\mu}^{EW} } \approx x \;
\left( M_W R \right)^2 S_{\delta} / ( \delta - 2 )
\left( M R \right)^{
\left( \delta - 2 \right)},
\label{amukk2}
\end{equation}
where $x$ is a factor of $O(1)$ 
which includes two effects as follows. One effect
is from the loop
integral. The reason is that in
the case of SM electroweak correction there
is only one heavy particle in the loop ($W$)
whereas in the case of the KK
contribution both the particles in the loop ($W_{KK}$
and $\nu_{KK}$) are heavy. Hence the ratio
of the contribution
from a KK state of mass $n / R$ to the SM electroweak contribution 
is not simply
the ratio of heaviest masses in the loop,
i.e., it is 
$\left( M_W / \left( n / R \right)
\right)^2$ {\em only} up to a factor of $O(1)$.
In principle, $x$ can be computed in the effective $4D$ theory. However,
we see from Eq. (\ref{amukk2}) that this factor can be absorbed into
the definition of the cut-off $M$.
The other effect included in the factor $x$ is that
$a_{\mu}^{EW}$ includes the $W$ and $Z$ loop contributions
in the SM, whereas $a_{\mu}^{KK}$ includes loop contributions from
KK states of $W$, $Z$ {\em and} the photon. 
 
There is an upper limit on $( M R )^{\delta}$
from the condition that the effective $4D$ theory at  
the cut-off
$M$
remain perturbative.
The loop expansion parameter at the cut-off $M$ is given by
$\sim N_{KK} N_i g_2^2 / \left( 16 \pi^2 \right)$. Here
$N_i$ is the number of ``colors'' and
$N_{KK} \approx S_{\delta} / \delta \left( M R \right)^{\delta}$ is 
the number of KK states
lighter than $M$, i.e, the total number of KK states in the 
effective $4D$ theory.
We cannot trust perturbative calculations 
if this factor 
is larger than $1$.
Therefore, we impose the constraint
\begin{equation}
N_{KK} \approx
S_{\delta} / \delta
\left( M R \right)^{\delta} \stackrel{<}{\sim} 16 \pi^2
\label{pertconstraint}
\end{equation}
(QCD effective coupling is smaller since number of KK states
of gluons and quarks is smaller). Thus, the upper limit on $M$ is not
much larger than $R^{-1}$. Nonetheless, we have checked if
$N_{KK} \gg 1$ (i.e., $M$ is close to the upper limit of Eq.
(\ref{pertconstraint})), then
the sum over KK states can {\em still}
be approximated by an integral.

\begin{figure}
\centerline{\epsfxsize=0.6\textwidth \epsfbox{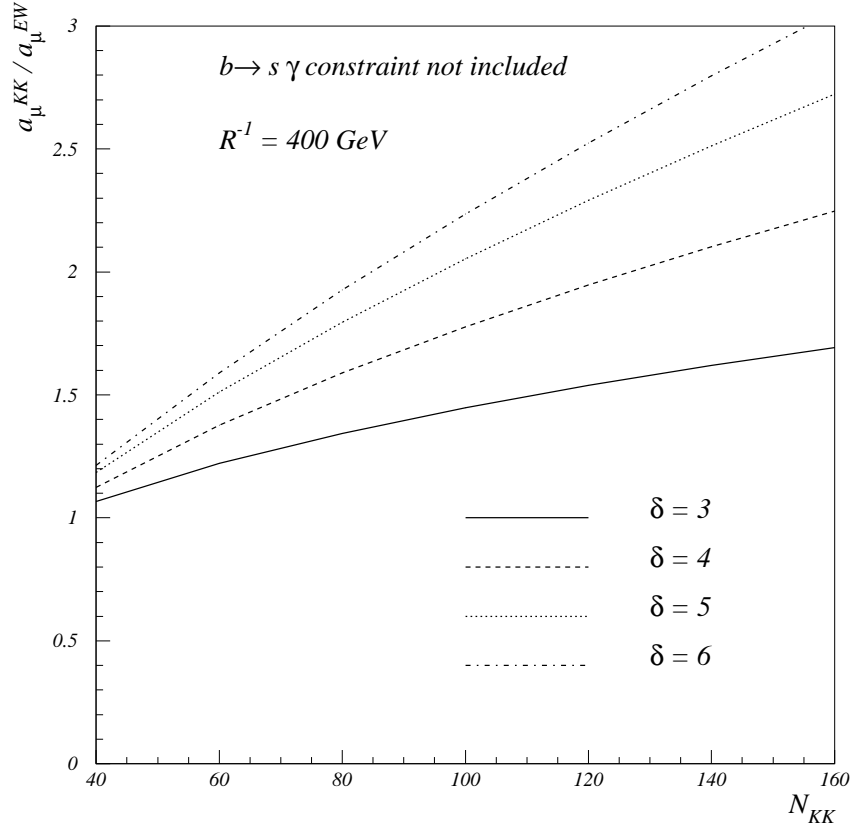}}
\caption{
The contribution to the MMM from KK states of the electroweak gauge
bosons relative to the SM electroweak correction 
as a function
of the total number of KK states (see Eq.
(\ref{amukk3})) and for $\delta = 3, 4, 5, $ and $6$, where
$\delta$ is the number of extra dimensions in which
the leptons and electroweak gauge bosons propagate. 
We assume $x =1$, $R^{-1} = 400$ GeV and do not
impose the constraint from the process $b \rightarrow s \gamma$
since this constraint depends sensitively on the Higgs content (see text).
}
\protect\label{mmmplot}
\end{figure}

We can rewrite Eq. (\ref{amukk2}) in terms of
$N_{KK}$ (using $N_{KK} \approx S_{\delta} / \delta
\left( MR \right) ^ {\delta}$)
as follows:
\begin{equation}
\frac{ a_{\mu}^{KK} }{ a_{\mu}^{EW} } \approx x \;
\left( M_W R \right)^2 S_{\delta} / ( \delta - 2 ) \;
\left( \delta / S_{\delta} N_{KK} \right)^{ ( \delta -2 ) / \delta }
\label{amukk3}
\end{equation}
In Fig. \ref{mmmplot}, we plot $a_{\mu}^{KK} / a_{\mu}^{EW}$ as a function
of $N_{KK}$ for different values of $\delta$ and for
$R^{-1} = 400$ GeV. From
Eq. (\ref{pertconstraint}),
the upper limit on $N_{KK}$ is about $16 \pi^2
\approx 160$.
We have chosen $x = 1$ for this plot; as mentioned
before, if $x \neq 1$, then it can be absorbed into the 
definition of the cut-off, or in other words, in the value of
$N_{KK}$ (see Eq. (\ref{amukk3})).
Also, in this figure, we have not imposed the constraint
from BR$\left(B \rightarrow X_s \gamma \right)$
(see the discussion later). 
We see from this figure that it is possible to
have $a_{\mu}^{KK} \sim a_{\mu}^{EW}$ for $\delta \geq 3$
as long as $N_{KK} \stackrel{>}{\sim} 40$.

We
now discuss the values of
$\delta _c$ for which the constraints from $T$ and hadronic
$B$ decays can be satisfied.

Since $\sum_n^{\delta} 1 / n^2 = \pi^2 / 3$ for $\delta = 1$,
it is clear from Eq. (\ref{B3b}) that if $\left( \delta - \delta _c \right)
= 1$, then $r_B \leq 13 \%$ provided $R^{-1} \geq 400$ GeV.
Thus, it is possible to (barely) satisfy the constraint from
hadronic $B$ decays by choosing $\delta _c = \delta -1$.

For given values of
$\delta$ and $\delta _c$, in general, it should be possible to choose
the value of $M_H$ such that
$T \approx 0$ (see Eq. (\ref{T3b}) or Eq. (\ref{Tconstraint})). 
\footnote{A caveat is that as mentioned earlier, if
$x \neq 1$, then the results in Fig.  \ref{mmmplot} really 
correspond to a ``redefined'' value
of $N_{KK}$ (or in other words, the cut-off $M$). Then, the cut-off
appearing in the sum over KK states in Eq. (\ref{T3b}) is, strictly 
speaking, different than that in Fig.  \ref{mmmplot}. Thus, 
for a given value of
$N_{KK}$ in Fig.  \ref{mmmplot} 
the value of $M_H$ which gives $T \approx 0$ is
modified.
However, the two
values of cut-off and hence the two values 
of $M_H$ differ by a factor of at most $O(1)$.}
Note that a small non-zero value of
$T$ is allowed (at $2 \sigma$ level, 
$- 0.3
\stackrel{<}{\sim} T \stackrel{<}{\sim} 0.4$ as mentioned before) and
similarly the constraint from
hadronic $B$ decays is also not very precise. Thus, for
given values of other parameters, there can be a  
{\em range} of values of $\delta _c$ and $M_H$ which is 
consistent with $T$ and with hadronic $B$ decays.

Next, we discuss the constraints
from
measurements of 
$\Gamma \left( b \rightarrow s \gamma \right) / \Gamma 
\left( b \rightarrow c l \nu \right)$. 
The semileptonic decay is not affected by the KK states of $W$ since these
states do not couple to (zero-mode) leptons.
The amplitude
for the process $b \rightarrow s \gamma$ gets a contribution from
KK states of $W$
$\sim g_2^2 / \left( 16 \pi^2 \right) m_b 
V^{\star}_{tb} V_{ts} m_t^2 \sum_n ^{\delta}
\left( n / R \right) ^4$ (the factor of $m_t^2$ reflects GIM cancelation).
This is to be compared to the SM contribution
$\sim g_2^2 / \left( 16 \pi^2 \right) m_b V^{\star}_{tb} V_{ts}
1 / M_W^2$: the ratio is   
$\sim \left( M_W m_t R^2 \right)^2 \sum ^{\delta}_n 1 / n^4$.
Thus, even if Eq. (\ref{amuconstraint}) is satisfied,
i.e., the KK contribution to
$a_{\mu}$ is comparable to the SM electroweak correction,
the $W_{KK}$ contribution to
$b \rightarrow s \gamma$ is smaller than in the SM.
This
is due to (a) an additional suppression by
a factor of 
$1 / \left( m_t R \right)^2 \stackrel{>}{\sim} 5$ (since $R^{-1}
\stackrel{>}{\sim} 400$ GeV)
in the case of $b \rightarrow s \gamma$ (as compared to $a_{\mu}$)
and (b) $\sum _n ^{\delta} 1 / n^4 < \sum _n ^{\delta} 1/ n^2$. 

However, there is a larger contribution to the amplitude
for $b \rightarrow s \gamma$ from loops with KK states of 
charged would-be-Goldstone boson (WGB) as follows. 
With only one Higgs doublet,
the {\em zero-mode} charged WGB is unphysical
since it becomes 
the longitudinal component of $W$. However, the 
{\em excited (KK)} states of charged WGB 
{\em are} physical and have masses $\sim n / R$
(assuming $1/R \gg M_H$). The coupling
of the KK states of
charged WGB to right-handed top quark is the same as for the zero-mode
charged WGB (i.e., longitudinal $W$) and is $\lambda _t \approx
\sqrt{2} m_t / v$. Thus, the loop diagram with KK states
of charged WGB and top quark gives an amplitude for
$b \rightarrow s \gamma$ given by 
$\sim \lambda _t ^2 / \left( 16 \pi^2 \right) m_b V^{\star}_{tb} V_{ts}
\sum _n ^{\delta}
\left( n / R \right)^2$. The ratio of this contribution
to the SM amplitude is
$\sim \left( m _t R \right)^2 \sum _n ^{\delta}
1/ n^2$
and is clearly larger than 
the $W_{KK}$ contribution.
Moreover,
if Eq. (\ref{amuconstraint}) is
satisfied, i.e., $a_{\mu}^{KK} \sim a_{\mu}^{EW}$,
then this additional amplitude for $b \rightarrow s \gamma$
is also comparable to the SM amplitude. 
Such a large contribution
to the process $b \rightarrow s \gamma$ is 
not allowed since the 
SM prediction for the rate 
agrees with  
experiment to within $\sim 20 \%$ (combining theory and experiment 
$1 \sigma$ errors).
Thus, with one Higgs doublet (and assuming no other new
physics contributions to $b \rightarrow s \gamma$), we see that
the constraint from 
$B \rightarrow X_s \gamma$
rules out the possibility that the contribution to the MMM from KK states of
electroweak gauge bosons is comparable
to the SM electroweak correction.

However, if the Higgs sector is {\em non-minimal}, 
then the contribution of KK states of charged Higgs
to the process
$b \rightarrow s \gamma$
is modified.  
For example, in the $2$-Higgs-doublet model (Model-II)
\cite{bsgamma}, 
a cancelation occurs between the contribution
of KK states of {\em physical} charged Higgs and the contribution of
the KK states of the charged WGB
mentioned above. 
This opens up the possibility
that $a_{\mu}^{KK} \sim a_{\mu}^{EW}$. Note that the
loop contribution to
$a_{\mu}$ from  
KK states of charged Higgs is negligible in this model
since it is suppressed
by additional powers of $m_{\mu}$ compared
to the contribution due to KK states of $W$.

\section{Conclusions}
In summary, we have revisited the contributions to the
anomalous magnetic moment of the muon (MMM) in models with
flat
extra dimensions of size $\sim$ TeV$^{-1}$ accessible to the SM particles.
In particular, we analyzed
a model with colored particles
(quarks and gluon) propagating in a sub-space
of the bulk in which non-colored particles (leptons,
Higgs and electroweak gauge bosons) propagate. In such a model,
the constraints
on the size of the extra dimensions from tree-level processes and
the $T$ parameter (which
have been analyzed previously) can be evaded. Thus, in
this model 
the contribution to the MMM from the KK states of electroweak gauge
bosons can be (potentially)
comparable to the SM electroweak correction
(unlike in other types of models).
However, with only one
Higgs doublet,
once the constraint from contribution of KK states 
to the process $b \rightarrow s \gamma$
(which has {\em not} been considered before) is imposed, we have shown 
that such 
a large contribution to the MMM is not possible.
The $b \rightarrow s \gamma$ 
constraint can be relaxed in models with more than
one Higgs doublet and then
a large contribution to the MMM becomes possible.

\end{document}